\documentclass[11pt,a4paper]{article}
\usepackage{amsmath}
\usepackage{amsfonts}
\usepackage{diagbox}
\usepackage{array}
\usepackage{bm}
\usepackage{tikz}
\usetikzlibrary{fit}					
\usetikzlibrary{backgrounds}	%

\usepackage{mathrsfs}
\usepackage{cite}

\begin{document}

\title{Higher $\varepsilon$-poles and logarithms in the MS-like schemes from the algebraic structure of the renormalization group}

\author{
N.P.Meshcheriakov\,${}^{a}$, V.V.Shatalova\,${}^{b}$, and K.V.Stepanyantz${}^{c}$ $\vphantom{\Big(}$
\medskip\\
${}^a${\small{\em Moscow State University, Faculty of Physics,}}\\
{\small{\em Department of Quantum Theory and High Energy Physics,}}\\
{\small{\em 119991, Moscow, Russia}}\\
\vphantom{1}\vspace*{-2mm}\\
${}^b${\small{\em Moscow State University, AESC MSU – Kolmogorov boarding school}},\\
{\small{\em Department of Physics,}}\\
{\small{\em 119192, Moscow, Russia}}\\
\vphantom{1}\vspace*{-2mm}\\
${}^c${\small{\em Moscow State University, Faculty of Physics,}}\\
{\small{\em Department of Theoretical Physics,}}\\
{\small{\em 119991, Moscow, Russia}}\\
\vphantom{1}\vspace*{-2mm}\\
}

\maketitle

\begin{abstract}
We investigate the structure of renormalization constants within the MS-like renormalization prescriptions for a version of dimensional regularization in which the dimensionful regularization parameter $\Lambda$ differs from the renormalization point $\mu$. Namely, we rewrite the all-loop equations relating coefficients at higher $\varepsilon$-poles and higher powers of $\ln\Lambda/\mu$ to the coefficients of the renormalization group functions in a simple unified form. It is argued that this form follows from the algebraic structure of the renormalization group.
\end{abstract}

\allowdisplaybreaks

\section{Introduction}
\hspace*{\parindent}

In most quantum field theory models quantum corrections are divergent in the ultraviolet region. For the renormalizable theories these divergences are removed by the renormalization procedure, certainly, after a proper regularization of a theory. In the case of using dimensional regularization \cite{'tHooft:1972fi,Bollini:1972ui,Ashmore:1972uj,Cicuta:1972jf} or reduction \cite{Siegel:1979wq} the renormalization constants contain $\varepsilon$-poles, where $\varepsilon\equiv 4-D$ (for theories in the four-dimensional space-time). It is well known that the coefficients at higher $\varepsilon$-poles are related to the coefficients of the renormalization group functions (RGFs), namely, of the $\beta$-function and the anomalous dimension(s) by the 't Hooft pole equations \cite{tHooft:1973mfk}, see \cite{Kazakov:2008tr} for a review. The generalizations of these equations to different theories can be found in \cite{Kazakov:1987jp,Kazakov:2016wrp,Borlakov:2016mwp,Kazakov:2019wce,Solodukhin:2020vuw}. The explicit solution of the 't Hooft equations for the charge renormalization constant in the minimal subtraction (MS) scheme can be found in \cite{Ivanov:2017ekx,Ivanov:2018pga}.

However, there are regularizations which are formulated in integer dimensions \cite{Gnendiger:2017pys}, for instance, the higher covariant derivative regularization \cite{Slavnov:1971aw,Slavnov:1972sq,Slavnov:1977zf}. In the supersymmetric case this regularization can be formulated in terms of superfields \cite{Krivoshchekov:1978xg,West:1985jx,Aleshin:2016yvj,Kazantsev:2017fdc} and does not break supersymmetry. Some other advantages of the higher covariant derivative method for regularizing supersymmetric theories are discussed, e.g., in \cite{Stepanyantz:2019lyo}. Therefore, the regularizations of this type are also worth considering. In the case of using them the divergences are regularized by introducing a scale $\Lambda$, which plays a role of an ultraviolet cut-off. If the divergences are logarithmic, then the renormalization constants contains powers of $\ln\Lambda/\mu$. This logarithm is analogous to $\varepsilon^{-1}$ in the case of using the dimensional technique for regularization. In particular, the coefficients at higher powers of $\ln\Lambda/\mu$ can also be expressed in terms of the coefficients of various RGFs, see \cite{Derkachev:2017nhd,Meshcheriakov:2022tyi}. However, the equations relating coefficients at higher powers of $\ln\Lambda/\mu$ in the case of using a cut-off type regularization to the coefficients of RGFs and similar equations for the coefficients at higher $\varepsilon$-poles are essentially different. It is convenient to establish the relations between these coefficients using a version of the dimensional technique in which the dimensionful regularization parameter $\Lambda$ is different from the renormalization point $\mu$. Application of this regularization to calculating multiloop quantum corrections can be found in, e.g., \cite{Aleshin:2015qqc,Aleshin:2016rrr,Aleshin:2019yqj}. In this case the renormalization constants contain higher $\varepsilon$-poles, higher powers of $\ln\Lambda/\mu$, and the mixed terms. The coefficients at all these terms can be expressed via RGFs with the help of the general formulas derived in \cite{Meshcheriakov:2023fmk}. The starting point of this derivation is the independence of the (four-dimensional) RGFs from $\varepsilon^{-1}$ and $\ln\Lambda/\mu$, which is actually a consequence of the renormalization group \cite{StueckelbergdeBreidenbach:1952pwl,Gell-Mann:1954yli,Bogolyubov1,Bogolyubov:1956gh}, see also \cite{Bogolyubov:1980nc,Collins:1984xc,Stevenson:2022gcv}. Therefore, one can guess \cite{Derkachev:2017nhd} that such equations can arise from the group theory. The structure of the renormalization group was discussed in \cite{Osborn:1991gm,Kovalev:1996ze,Shirkov:1999hj,Kovalev:2008ht,Kamenshchik:2020yyn}. Its generators and their commutation relations have been constructed in \cite{Kataev:2024xbl}. In particular, the Lie algebra corresponding to the finite renormalizations of charge appears to be a certain subalgebra of the Witt algebra \cite{Cartan} (whose central extension is the widely known Virasoro algebra \cite{Virasoro:1969zu}). It is important that the group of finite renormalizations also contains a subgroup corresponding to the changes of the renormalization point, which is Abelian \cite{Brodsky:1992pq,Brodsky:2012ms}. Using the algebraic structure of this subgroup in this paper we rewrite the general equations derived in \cite{Meshcheriakov:2023fmk} for the MS-like subtraction schemes in such a form that their group theory origin becomes clear.

The paper is organized as follows. In Sect. \ref{Section_Regularization} we recall how it is possible to formulate dimensional regularization in the case when the dimensionful regularization parameter $\Lambda$ differs from the renormalization point $\mu$. Next, in Sect. \ref{Section_Renormalization_Constants} we present the equations that relate the coefficients at higher $\varepsilon$-poles, higher powers of $\ln\Lambda/\mu$, and the mixed terms in various renormalization constants to the coefficients of RGFs derived in \cite{Meshcheriakov:2023fmk} and rewrite them in a different form. This new form is used for revealing the renormalization group origin of the equations under consideration in Sect. \ref{Section_Renormalization_Group_Origin}. In particular, it is demonstrated that they can be obtained with the help of the exponential map for the rescaling subgroup of the renormalization group. The results are summarized and briefly discussed in Conclusion. Appendices contain some technical details of the calculations.

\section{Dimensional regularization with $\Lambda\ne \mu$}
\hspace*{\parindent}\label{Section_Regularization}

In the dimension $D\ne 4$ the gauge coupling constant $\widetilde\alpha_0$ has the dimension $m^\varepsilon$ and can, therefore, be presented as $\widetilde\alpha_0 = \alpha_0\Lambda^\varepsilon$, where $\Lambda$ is a parameter with the dimension of mass. Then the renormalization of the coupling constant can be made according to the prescription

\begin{equation}\label{Alpha_D_Renormalization}
\alpha_0 = \Big(\frac{\mu}{\Lambda}\Big)^{\varepsilon} \bm{\alpha}\, \bm{Z_\alpha}^{-1}(\bm{\alpha}, \varepsilon^{-1}),
\end{equation}

\noindent
where $\mu$ is a renormalization point and $\bm{\alpha}$ is the renormalized gauge coupling. In the case of using the background field method \cite{DeWitt:1965jb,Arefeva:1974jv,Abbott:1980hw,Abbott:1981ke} the charge renormalization constant can be obtained by calculating the divergent part of the two-point Green's function of the background gauge field. In the case of using the $\mbox{MS}$ scheme \cite{tHooft:1973mfk} the renormalization constant includes only $\varepsilon$-poles. The modified minimal subtraction ($\overline{\mbox{MS}}$) \cite{Bardeen:1978yd} is obtained after the redefinition of the renormalization point

\begin{equation}\label{MS_Bar_Mu}
\mu \to \frac{\mu\,\exp(\gamma/2)}{\sqrt{4\pi}},
\end{equation}

\noindent
where $\gamma\equiv - \Gamma'(1)\approx 0.577$.

Field renormalization constant $\bm{Z}(\bm{\alpha},\varepsilon^{-1})$ is defined by requiring the finiteness of the corresponding renormalized Green's function $G_R$ in the limit $\varepsilon\to 0$,

\begin{equation}\label{G_D_Renormalization}
G_R\Big(\bm{\alpha},\ln\frac{\mu}{P}\Big) = \lim\limits_{\varepsilon\to 0}\bm{Z}(\bm{\alpha},\varepsilon^{-1})\, G\Big[\Big(\frac{\mu}{P}\Big)^\varepsilon\bm{\alpha} \bm{Z_\alpha}^{-1}(\bm{\alpha},\varepsilon^{-1}),\varepsilon^{-1}\Big],
\end{equation}

\noindent
where $P$ is the external momentum, and, for simplicity, we consider a two-point Green's function. However, it is much more convenient to deal with RGFs which encode ultraviolet divergences in the most concise form. For the above described way of renormalization the $\beta$-function and the anomalous dimension are defined by the equations

\begin{equation}\label{RGFs_D_Definition}
\bm{\beta}(\bm{\alpha},\varepsilon) \equiv \frac{d\bm{\alpha}(\alpha_0(\Lambda/\mu)^\varepsilon,\varepsilon^{-1})}{d\ln\mu}\bigg|_{\alpha_0=\text{const}};\qquad
\bm{\gamma}(\bm{\alpha}) \equiv \frac{d\ln \bm{Z}(\bm{\alpha}, \varepsilon^{-1})}{d\ln\mu} \bigg|_{\alpha_0=\text{const}} = \bm{\beta}(\bm{\alpha},\varepsilon) \frac{\partial \ln \bm{Z}}{\partial \bm{\alpha}}.
\end{equation}

Alternatively, the renormalization can be made in the four-dimensional form

\begin{eqnarray}\label{Alpha_4_Renormalization}
&& \frac{1}{\alpha_0} = \frac{Z_\alpha (\alpha, \varepsilon^{-1}, \ln \Lambda/\mu)}{\alpha};\\
\label{G_4_Renormalization}
&& G_R\Big(\alpha,\ln\frac{\mu}{P}\Big) = \lim\limits_{\varepsilon\to 0} Z(\alpha,\varepsilon^{-1},\ln\Lambda/\mu)\, G\Big[\Big(\frac{\Lambda}{P}\Big)^\varepsilon\alpha Z_\alpha^{-1}(\alpha,\varepsilon^{-1},\ln\Lambda/\mu), \varepsilon^{-1}\Big].\qquad
\end{eqnarray}

\noindent
In this case the renormalization constants $Z_\alpha$ and $Z$ should not contain positive powers of $\varepsilon$ (which are very essential in Eqs. (\ref{Alpha_D_Renormalization}) and (\ref{G_D_Renormalization})). Therefore, (for the same bare coupling $\alpha_0$) the renormalized coupling $\bm{\alpha}$ differs from the renormalized coupling $\alpha$. That is why we denote the former one in bold. For the four-dimensional renormalization procedure defined by Eqs. (\ref{Alpha_4_Renormalization}) and (\ref{G_4_Renormalization}) RGFs are defined as

\begin{equation}\label{RGFs_4_Definition}
\beta(\alpha) \equiv \frac{d \alpha(\alpha_0, \varepsilon^{-1}, \ln \Lambda/\mu) }{d \ln \mu}\bigg|_{\alpha_0=\text{const}};\qquad
\gamma(\alpha) \equiv \frac{d\ln Z(\alpha,\varepsilon^{-1},\ln\Lambda/\mu)}{d\ln\mu}\bigg|_{\alpha_0=\text{const}}.
\end{equation}

It is known (see, e.g., \cite{Kazakov:2008tr,Meshcheriakov:2023fmk}) that the $D$-dimensional RGFs (\ref{RGFs_D_Definition}) are related to the four-dimensional ones (\ref{RGFs_4_Definition}) by the equations

\begin{equation}\label{RGFs_Relations}
\bm{\beta}(\alpha,\varepsilon) = - \varepsilon\alpha + \beta(\alpha);\qquad \bm{\gamma}(\alpha) = \gamma(\alpha).
\end{equation}

\noindent
It is important that (unlike the corresponding renormalization constants) the functions $\beta(\alpha)$ and $\gamma(\alpha)$ do not depend on $\varepsilon$ and $\ln\Lambda/\mu$.

\section{Expressions for the renormalization constants in the $\mbox{MS}$-like schemes}
\hspace*{\parindent}\label{Section_Renormalization_Constants}

It was noticed a long time ago \cite{tHooft:1973mfk} that the coefficients at higher $\varepsilon$-poles are related to the coefficients at simple ($\varepsilon^{-1}$) poles, which determine RGFs. So does the coefficients at  powers of $\ln\Lambda/\mu$, which are also related to the coefficients of RGFs, see, e.g., \cite{Collins:1984xc}. For the regularization considered in this paper the corresponding equations can be written in the form of the following all-order exact expressions derived in \cite{Meshcheriakov:2023fmk},

\begin{eqnarray}\label{LnZ_Original_Result}
&&\hspace*{-9mm} \frac{\partial\ln Z_\alpha}{\partial\ln\alpha} = 1 - \exp\Big\{ \ln\frac{\Lambda}{\mu}\, \frac{\hat\partial}{\partial\ln\alpha} \frac{\beta(\alpha)}{\alpha}\Big\}\, \Big(1-\frac{\beta(\alpha)}{\varepsilon\alpha}\Big)^{-1};\\
\label{ZS_Original_Result}
&&\hspace*{-9mm} \Big(\frac{\partial}{\partial\ln\alpha} -S\Big) (Z_\alpha)^S = - S \exp\Big\{ \ln\frac{\Lambda}{\mu}\, \Big(\frac{\hat\partial}{\partial\ln\alpha} - S\Big) \frac{\beta(\alpha)}{\alpha}\Big\}\, \Big(1-\frac{\beta(\alpha)}{\varepsilon\alpha }+ S \int\limits^{\wedge} \frac{d\alpha}{\alpha}\, \frac{\beta(\alpha)}{\varepsilon\alpha} \Big)^{-1};\\
\label{LnZM_Original_Result}
&&\hspace*{-9mm} \frac{\partial\ln Z}{\partial\ln\alpha} = \frac{\alpha\gamma(\alpha)}{\beta(\alpha)} - \exp\Big\{ \ln\frac{\Lambda}{\mu}\, \frac{\hat\partial}{\partial\ln\alpha} \frac{\beta(\alpha)}{\alpha}\Big\}\,
\bigg[\frac{\alpha\gamma(\alpha)}{\beta(\alpha)} \Big(1-\frac{\beta(\alpha)}{\varepsilon\alpha}\Big)^{-1}\bigg].\vphantom{\int\limits^{\wedge}}
\end{eqnarray}

\noindent
Here it is assumed that all expressions are considered as their formal series expansions in $\alpha$, and a hat over the operator $\partial/\partial\ln\alpha$ means that this operator acts on everything to the right of it. Similarly, the integral operator with a hat acts on everything to the right of it. Note that in the above equations it can act only on polynomials in $\alpha$, 

\begin{equation}
\int\limits^{\wedge} d\alpha\,\alpha^n \equiv \frac{\alpha^{n+1}}{n+1},\qquad n\ge 0,
\end{equation}

\noindent
so that the integration is taken from $0$ to $\alpha$. The explicit expressions for the renormalization constants in the lowest loops can easily be obtained by expanding the expressions (\ref{LnZ_Original_Result}) --- (\ref{LnZM_Original_Result}) in powers of $\alpha$. Up to the five loops (and six loops for $Z_\alpha$) they are explicitly written and verified in \cite{Meshcheriakov:2023fmk}. From Eqs. (\ref{LnZ_Original_Result}) --- (\ref{LnZM_Original_Result}) it is also possible to find the relation between the coefficients at pure higher poles and pure higher logarithms. (For example, the lowest poles and logarithms in the $\mbox{MS}$-like schemes always enter the renormalization constants in the combination $1/L\varepsilon + \ln\Lambda/\mu$, where $L$ is a number of loops, \cite{Chetyrkin:1980sa}.)

The main idea underlying the derivation of Eqs. (\ref{LnZ_Original_Result}) --- (\ref{LnZM_Original_Result}) is that the (four-dimensional) $\beta$-function and the anomalous dimension depend only on $\alpha$ and are independent of $\varepsilon$ and $\ln\Lambda/\mu$. This fact, in turn, is closely related to the group structure of renormalization \cite{Bogolyubov:1980nc,Collins:1984xc}. Therefore, we expect that the above equations can be derived directly from the renormalization group considerations. To establish this correspondence, we first rewrite Eqs. (\ref{LnZ_Original_Result}) --- (\ref{LnZM_Original_Result}) in the equivalent form

\begin{eqnarray}\label{LnZ_Result}
&& \ln Z_\alpha = \ln\alpha - \exp\Big(\ln\frac{\Lambda}{\mu}\, \beta(\alpha)\frac{\partial}{\partial\alpha} \Big)\,\bigg\{\int\limits_0^\alpha \frac{d\alpha}{\alpha}
\bigg[\Big(1-\frac{\beta(\alpha)}{\varepsilon\alpha}\Big)^{-1}-1\bigg] +\ln\alpha\bigg\};\\
\label{ZS_Result}
&& (Z_\alpha)^S = \alpha^S \exp\Big(\ln\frac{\Lambda}{\mu}\,\beta(\alpha) \frac{\partial}{\partial\alpha}\Big)\,\alpha^{-S} \exp\bigg\{-S\int\limits_0^\alpha\frac{d\alpha}{\alpha}\bigg[\Big(1-\frac{\beta(\alpha)}{\varepsilon\alpha}\Big)^{-1}-1\bigg]\bigg\};\\
\label{LnZM_Result}
&& \ln Z = \int\limits_a^\alpha d\alpha\,\frac{\gamma(\alpha)}{\beta(\alpha)} +  \exp\Big(\ln\frac{\Lambda}{\mu}\, \beta(\alpha) \frac{\partial}{\partial\alpha}\Big) \bigg[\int\limits_0^\alpha d\alpha\, \frac{\gamma(\alpha)}{\beta(\alpha)-\varepsilon\alpha} - \int\limits_a^\alpha d\alpha\,\frac{\gamma(\alpha)}{\beta(\alpha)}\bigg],\qquad
\end{eqnarray}

\noindent
where the constant $a$ at the lower limits of two integrals in the last equation can be arbitrary.\footnote{Note that here, for simplicity, we omit hats over $\partial/\partial\alpha$, because these expressions are unambiguous.} Deriving these expressions we also took into account that for $\alpha=0$ the renormalization constants $Z_\alpha$ and $Z$ should be equal to 1. Details of the calculation can be found in Appendix \ref{Appendix_Section_Rewriting}. Eqs. (\ref{LnZ_Result}) --- (\ref{LnZM_Result}) can also be presented as

\begin{eqnarray}\label{LnZ_Result2}
&&\hspace*{-7mm} \ln \alpha_0 = \ln(\alpha Z_\alpha^{-1}) = \exp\Big(\ln\frac{\Lambda}{\mu}\,\beta(\alpha)\frac{\partial}{\partial\alpha}\Big)\,
\bigg\{\int\limits_0^\alpha \frac{d\alpha}{\alpha}
\bigg[\Big(1-\frac{\beta(\alpha)}{\varepsilon\alpha}\Big)^{-1}-1\bigg] +\ln\alpha\bigg\};\\
\label{ZS_Result2}
&&\hspace*{-7mm} \alpha_0^{-S} = \alpha^{-S} (Z_\alpha)^S = \exp\Big(\ln\frac{\Lambda}{\mu}\,\beta(\alpha) \frac{\partial}{\partial\alpha}\Big)\,\alpha^{-S} \exp\bigg\{-S\int\limits_0^\alpha\frac{d\alpha}{\alpha}\bigg[\Big(1-\frac{\beta(\alpha)}{\varepsilon\alpha}\Big)^{-1}-1\bigg]\bigg\};\\
\label{LnZM_Result2}
&&\hspace*{-7mm} \ln Z(\alpha,\varepsilon^{-1},\ln\Lambda/\mu) - \int\limits_a^\alpha d\alpha\,\frac{\gamma(\alpha)}{\beta(\alpha)} = \exp\Big(\ln\frac{\Lambda}{\mu}\,\beta(\alpha)\frac{\partial}{\partial\alpha}\Big)
\bigg[ \ln Z(\alpha,\varepsilon^{-1},0) - \int\limits_a^\alpha d\alpha\,\frac{\gamma(\alpha)}{\beta(\alpha)} \bigg].\quad
\end{eqnarray}

\noindent
Note that writing the last equation we took into account that the expression $\ln Z(\alpha,\varepsilon^{-1},0)$ coincides with $\ln\bm{Z}(\alpha,\varepsilon^{-1})$ and, according to Eqs. (\ref{RGFs_D_Definition}) and (\ref{RGFs_Relations}), is given by

\begin{equation}\label{LnZM_Pure_Poles}
\ln Z(\alpha,\varepsilon^{-1},0) = \ln\bm{Z}(\alpha,\varepsilon^{-1}) = \int\limits_0^\alpha d\alpha\, \frac{\gamma(\alpha)}{\beta(\alpha)-\varepsilon\alpha} = -\int\limits_0^\alpha d\alpha\, \frac{\gamma(\alpha)}{\beta(\alpha)} \Big[\Big(1-\frac{\beta(\alpha)}{\varepsilon\alpha}\Big)^{-1}-1\Big].
\end{equation}

\section{Renormalization group origin of the equations for the renormalization constants}
\hspace*{\parindent}\label{Section_Renormalization_Group_Origin}

According to \cite{Kataev:2024xbl}, the finite renormalizations corresponding to the changes of the renormalization scale

\begin{equation}\label{Rescaling}
\alpha(\mu) \to \alpha'(\mu) \equiv \alpha(\mu')
\end{equation}

\noindent
are generated by the operator

\begin{equation}
\hat L \equiv \beta(\alpha) \frac{d}{d\alpha},
\end{equation}

\noindent
so that the infinitesimal change of an arbitrary smooth function $f(\alpha)$ can be presented as

\begin{equation}
\delta f(\alpha) = \ln\frac{\mu'}{\mu}\, \beta(\alpha) \frac{d}{d\alpha} f(\alpha) = \ln\frac{\mu'}{\mu}\, \hat L f(\alpha),
\end{equation}

\noindent
where $\delta f$, by definition, includes only terms linear in $\ln\mu'/\mu$ in the expression $f(\alpha')-f(\alpha)$. Taking into account that the considered transformations form an Abelian subgroup of the renormalization group it is possible to obtain the corresponding finite transformations with the help of the exponential map \cite{Groote:2001im,Mikhailov:2004iq,Kataev:2014jba},

\begin{equation}\label{F_Finite_Shift}
f(\alpha') = \exp\Big(\ln\frac{\mu'}{\mu}\, \beta(\alpha)\frac{\partial}{\partial\alpha}\Big) f(\alpha).
\end{equation}

\noindent
Note that here we replace the total derivative $d/d\alpha$ by the partial derivative $\partial/\partial\alpha$ in order to stress that it does not act on $\ln\mu'/\mu$.

Let us first consider a simple case, when a theory is regularized by higher derivatives. Then, a renormalization prescription analogous to minimal subtraction is the HD+MSL scheme \cite{Kataev:2013eta,Shakhmanov:2017wji,Stepanyantz:2017sqg}, in which the renormalization constants include only powers of $\ln\Lambda/\mu$. This in particular implies that the function $\alpha(\alpha_0,\ln\Lambda/\mu)$ satisfies the boundary condition \cite{Kataev:2013eta}

\begin{equation}\label{Boundary_Condition}
\alpha(\alpha_0,\ln\Lambda/\mu)\Big|_{\mu=\Lambda} = \alpha_0,
\end{equation}

\noindent
so that all finite constants which determine the subtraction scheme are set to 0.

Choosing $\mu'=\Lambda$ in Eq. (\ref{Rescaling}) and taking into account that $\alpha'(\mu) = \alpha(\Lambda) = \alpha(\alpha_0,0)=\alpha_0$ from Eqs. (\ref{F_Finite_Shift}) and (\ref{Boundary_Condition}) for an arbitrary function $f(\alpha)$ in the HD+MSL scheme we obtain

\begin{equation}\label{Arbitrary_F}
f(\alpha_0) = \exp\Big(\ln\frac{\Lambda}{\mu}\, \beta(\alpha)\frac{\partial}{\partial\alpha}\Big) f(\alpha).
\end{equation}

\noindent
If we remove $\varepsilon$-poles formally taking the limit $\varepsilon^{-1}\to 0$ in Eqs. (\ref{LnZ_Result2}) and (\ref{ZS_Result2}), then the results coincide with Eq. (\ref{Arbitrary_F}) for $f(\alpha_0)=\ln\alpha_0$ and $f(\alpha_0) = \alpha_0^{-S}$, respectively. Taking into account the resemblance between the considered version of dimensional regularization and the regularizations of the cut-off type it is possible to note that the results in the latter case can simply be obtained by removing the $\varepsilon$-poles. Therefore, in, e.g., HD+MSL scheme the coefficients at higher logarithms in the renormalization constant $Z_\alpha$ are related to the coefficients of the $\beta$-function by Eq. (\ref{Arbitrary_F}) in the particular case $f(\alpha_0) = 1/\alpha_0$.

Similarly, the relations between coefficients at higher logarithms in the expression $\ln Z$ (where $Z$ is an arbitrary renormalization constant) and the coefficients of RGFs in the HD+MSL scheme can also be obtained from Eq. (\ref{Arbitrary_F}). For this purpose we note that the solution of the renormalization group equation for $\ln Z$ which satisfies the HD+MSL boundary conditions $\ln Z(\alpha_0,\ln\Lambda/\mu=0)=0$ and $\alpha(\alpha_0,\ln\Lambda/\mu=0)=\alpha_0$ is given by

\begin{equation}
\ln Z = \int\limits_{\alpha_0}^\alpha d\alpha\, \frac{\gamma(\alpha)}{\beta(\alpha)}.
\end{equation}

\noindent
Then, after removing $\varepsilon$-poles, Eq. (\ref{LnZM_Result}) can be presented in the form

\begin{equation}\label{LnZ_Preliminary}
\int\limits_a^{\alpha_0} d\alpha\,\frac{\gamma(\alpha)}{\beta(\alpha)} = \exp\Big(\ln\frac{\Lambda}{\mu}\,\beta(\alpha)\frac{\partial}{\partial\alpha}\Big)\, \int\limits_a^\alpha d\alpha\,\frac{\gamma(\alpha)}{\beta(\alpha)},
\end{equation}

\noindent
where $a$ is an arbitrary constant. We see that this equation exactly coincides with Eq. (\ref{Arbitrary_F}) in which the function $f$ is given by the expression

\begin{equation}
f(\alpha) = \int\limits_a^\alpha d\alpha\, \frac{\beta(\alpha)}{\gamma(\alpha)}.
\end{equation}

\noindent
Therefore, the HD+MSL equations for higher logarithms really follows from Eq. (\ref{Arbitrary_F}), which has the evident renormalization group origin.

Let us now consider a more complicated case of the regularization described in Sect. \ref{Section_Regularization}. In this version of dimensional regularization the renormalization constants contain not only $\varepsilon$-poles, but also powers of $\ln\Lambda/\mu$ and various mixed terms. The terms containing only powers of $\ln\Lambda/\mu$ have the same structure as the renormalization constants in the HD+MSL scheme, so that the regularization under consideration can be used for establishing the correspondence between the results obtained with the regularizations of the cut-off type and the regularizations involving the dimensional technique.

Setting $\mu=\Lambda$ in Eq. (\ref{LnZ_Result2}) we obtain the dependence of the renormalized coupling constant $\alpha$ on $\alpha_0$ and $\varepsilon^{-1}$ for the standard version of dimensional regularization,

\begin{equation}\label{Preliminary_Equality}
\alpha\exp\bigg\{\int\limits_0^\alpha\frac{d\alpha}{\alpha}\bigg[\Big(1-\frac{\beta(\alpha)}{\varepsilon\alpha}\Big)^{-1}-1\bigg]\bigg\}\bigg|_{\mu=\Lambda} = \alpha_0.
\end{equation}

\noindent
Note that the left hand side of Eq. (\ref{Preliminary_Equality}) coincide with the renormalization group invariant

\begin{equation}\label{Invariant1}
I_1\equiv \Big(\frac{\mu}{\Lambda}\Big)^\varepsilon \bm{\alpha}\exp\bigg\{\int\limits_0^{\bm{\alpha}}\frac{d\alpha}{\alpha}\bigg[\Big(1-\frac{\beta(\alpha)}{\varepsilon\alpha}\Big)^{-1}-1\bigg]\bigg\}
\end{equation}

\noindent
(where $\bm{\alpha}$ is the $D$-dimensional renormalized coupling constant defined by Eq. (\ref{Alpha_D_Renormalization})), which satisfies the equation

\begin{equation}\label{I1_Invariance}
\frac{d}{d\ln\mu} I_1 = 0.
\end{equation}

\noindent
proved in Appendix \ref{Appendix_Invariants}. Calculating the expression $I_1$ at $\mu=\Lambda$ and taking into account that for $\mu=\Lambda$ the coupling constants $\bm{\alpha}$ and $\alpha$ defined by Eqs. (\ref{Alpha_D_Renormalization}) and (\ref{Alpha_4_Renormalization}), respectively, coincide, we see that the left hand side of Eq. (\ref{Preliminary_Equality}) is really equal to $I_1$. (Evidently, $\alpha_0$ in the right hand of Eq. (\ref{Preliminary_Equality}) is also the renormalization group invariant.)

According to Eq. (\ref{F_Finite_Shift}) the expression in the left hand side of Eq. (\ref{Preliminary_Equality}) can be presented as

\begin{eqnarray}
&& \alpha\exp\bigg\{\int\limits_0^\alpha\frac{d\alpha}{\alpha}\bigg[\Big(1-\frac{\beta(\alpha)}{\varepsilon\alpha}\Big)^{-1}-1\bigg]\bigg\}\Bigg|_{\mu=\Lambda}
\nonumber\\
&&\qquad\qquad\qquad = \exp\Big(\ln\frac{\Lambda}{\mu}\, \beta(\alpha)\frac{\partial}{\partial\alpha}\Big) \Bigg(\alpha\exp\bigg\{\int\limits_0^\alpha\frac{d\alpha}{\alpha}\bigg[\Big(1-\frac{\beta(\alpha)}{\varepsilon\alpha}\Big)^{-1}-1\bigg]\bigg\}\Bigg).\qquad
\end{eqnarray}

\noindent
With the help of Eq. (\ref{Preliminary_Equality}) from this equation we obtain that for an arbitrary $\mu$ the bare coupling constant $\alpha_0$ can be expressed in terms of the four-dimensional renormalized coupling constant $\alpha$ as

\begin{equation}
\alpha_0 = \exp\Big(\ln\frac{\Lambda}{\mu}\, \beta(\alpha)\frac{\partial}{\partial\alpha}\Big) \Bigg(\alpha\exp\bigg\{\int\limits_0^\alpha\frac{d\alpha}{\alpha}\bigg[\Big(1-\frac{\beta(\alpha)}{\varepsilon\alpha}\Big)^{-1}-1\bigg]\bigg\}\Bigg).
\end{equation}

\noindent
Evidently, the group theory arguments used for deriving it are also valid for an arbitrary function $f$, so that

\begin{equation}\label{F_With_Exponent}
f(\alpha_0) = \exp\Big(\ln\frac{\Lambda}{\mu}\, \beta(\alpha)\frac{\partial}{\partial\alpha}\Big) f \Bigg(\alpha\exp\bigg\{\int\limits_0^\alpha\frac{d\alpha}{\alpha}\bigg[\Big(1-\frac{\beta(\alpha)}{\varepsilon\alpha}\Big)^{-1}-1\bigg]\bigg\}\Bigg).
\end{equation}

\noindent
In the particular cases $f(\alpha_0)=\ln\alpha_0$ and $f(\alpha_0) = \alpha_0^{-S}$ from this equation we obtain Eqs. (\ref{LnZ_Result2}) and (\ref{ZS_Result2}), respectively. Note that, as demonstrated in \cite{Meshcheriakov:2023fmk}, Eqs. (\ref{LnZ_Result2}) and (\ref{ZS_Result2}) encode all equations relating the coefficients at the higher $\varepsilon$-poles, logarithms, and mixed terms to the coefficients of the $\beta$-function. Therefore, these equations  really follow from the algebraic structure of the renormalization group.

Note that setting $\mu=\Lambda$ and $S=1$ in Eq. (\ref{ZS_Result}) we obtain that the renormalization constant $\bm{Z}_\alpha$ is given by the expression

\begin{equation}\label{Standard_Z}
\bm{Z}_\alpha(\alpha,\varepsilon^{-1}) = Z_\alpha(\alpha,\varepsilon^{-1},0) = \exp\bigg(\int\limits_0^\alpha\frac{d\alpha}{\alpha} \frac{\beta(\alpha)}{\beta(\alpha)-\varepsilon\alpha}\bigg)
\end{equation}

\noindent
(which, up to notations, agrees with \cite{Ivanov:2017ekx}). From Eq. (\ref{F_With_Exponent}) (for the function $f(\alpha_0)=1/\alpha_0$) we see that for an arbitrary value of $\ln\Lambda/\mu$ the renormalization constant $Z_\alpha$ can be written as

\begin{equation}\label{Z_Alpha_Preliminary}
Z_\alpha(\alpha,\varepsilon^{-1},\ln\Lambda/\mu) = \alpha \exp\Big(\ln\frac{\Lambda}{\mu}\, \beta(\alpha) \frac{\partial}{\partial\alpha}\Big) \Big(\alpha^{-1} Z_\alpha(\alpha,\varepsilon^{-1},0)\Big).
\end{equation}

Let us reveal how this equation is related to the algebraic structure of the renormalization group. For this purpose, first, we note that if the operators $A$ and $B$ satisfy the condition $[[A,B],B] = 0$, then

\begin{equation}\label{Algebraic_Identity}
e^{B} e^A e^{-B}  = \exp(e^{B} A e^{-B}) = \exp\Big(A - [A,B]\Big).
\end{equation}

\noindent
This identity can be applied for transforming Eq. (\ref{Z_Alpha_Preliminary}) if we take the operators

\begin{equation}
A \to \ln\frac{\Lambda}{\mu}\,\beta(\alpha)\frac{\partial}{\partial\alpha};\qquad B \to \ln\alpha,
\end{equation}

\noindent
for which the equation $[[A,B],B] = 0$ is evidently valid. As a result, the constant $Z_\alpha$ can be cast in the form

\begin{eqnarray}\label{Z_Alpha_Final}
&& Z_\alpha(\alpha,\varepsilon^{-1},\ln\Lambda/\mu) = \exp\Big\{\ln\frac{\Lambda}{\mu}\, \Big[\beta(\alpha) \frac{\partial}{\partial\alpha} - \frac{\beta(\alpha)}{\alpha}\Big]\Big\}  Z_\alpha(\alpha,\varepsilon^{-1},0)
\nonumber\\
&&\qquad\qquad\qquad\qquad\qquad
= \exp\Big\{\ln\frac{\Lambda}{\mu}\, \Big[\beta(\alpha) \frac{\partial}{\partial\alpha} - \gamma_\alpha(\alpha)\Big]\Big\}  \exp\bigg(\int\limits_0^\alpha\frac{d\alpha}{\alpha} \frac{\beta(\alpha)}{\beta(\alpha)-\varepsilon\alpha}\bigg),\qquad
\end{eqnarray}

\noindent
because the anomalous dimension of the coupling constant is

\begin{equation}\label{Gamma_Alpha}
\gamma_\alpha(\alpha) \equiv \frac{d\ln Z_\alpha}{d\ln\mu}\bigg|_{\alpha_0=\text{const}} = \frac{1}{\alpha_0 Z_\alpha} \frac{d\alpha}{d\ln\mu}\bigg|_{\alpha_0=\text{const}} = \frac{\beta(\alpha)}{\alpha}.
\end{equation}

\noindent
Expanding the expression (\ref{Z_Alpha_Final}) into a series in powers of $\alpha$ one obtains all equations relating the coefficients at higher powers of $\varepsilon$-poles, logarithms, and mixed terms to the coefficients of the $\beta$-function. However, in this paper we pay especial attention to the renormalization group origin of Eq. (\ref{Z_Alpha_Final}). To see it, we consider the rescaling transformations

\begin{equation}\label{Rescaling_Total}
Z(\alpha,\varepsilon^{-1},\ln\Lambda/\mu) \to Z'(\alpha,\varepsilon^{-1},\ln\Lambda/\mu) \equiv Z(\alpha,\varepsilon^{-1},\ln\Lambda/\mu'),
\end{equation}

\noindent
where $Z$ is an arbitrary renormalization constant. If the parameter $t\equiv \ln\mu'/\mu$ is small, then these transformations can be written in the infinitesimal form

\begin{equation}
\delta Z = t\, \frac{\partial Z}{\partial\ln\mu} = t\Big(\frac{d Z}{d\ln\mu} - \beta(\alpha)\frac{\partial Z}{\partial\alpha}\Big) = t\Big(\gamma(\alpha) -\beta(\alpha)\frac{\partial}{\partial\alpha}\Big) Z,
\end{equation}

\noindent
where the partial derivative $\partial/\partial\ln\mu$ acts only on explicitly written $\ln\mu$ (or, equivalently, is taken at $\alpha=\mbox{const}$), while the total derivative $d/d\ln\mu$ also acts on $\ln\mu$ inside $\alpha$. If we take into account that the rescaling transformations form an Abelian subgroup of the renormalization group, then the corresponding finite transformations should be obtained with the help of the exponential map, namely,

\begin{equation}\label{Finite_Rescalings}
Z(\alpha,\varepsilon^{-1},\ln\Lambda/\mu') = \exp\Big\{t\Big[\gamma(\alpha) -\beta(\alpha)\frac{\partial}{\partial\alpha}\Big]\Big\}\, Z(\alpha,\varepsilon^{-1},\ln\Lambda/\mu).
\end{equation}

\noindent
Taking $\mu'=\Lambda$ we can relate the renormalization constant at an arbitrary value of $\ln\Lambda/\mu$ to its value at $\ln\Lambda/\mu=0$,

\begin{equation}\label{Z_Rescaling}
Z(\alpha,\varepsilon^{-1},\ln\Lambda/\mu) = \exp\Big\{\ln\frac{\Lambda}{\mu}\Big[\beta(\alpha)\frac{\partial}{\partial\alpha} - \gamma(\alpha)\Big]\Big\}\, Z(\alpha,\varepsilon^{-1},0).
\end{equation}

\noindent
In the case under consideration $Z \to Z_\alpha$ and $\gamma \to \gamma_\alpha = \beta/\alpha$, so that from this equation we immediately reproduce Eq. (\ref{Z_Alpha_Final}).

In fact, an equation similar to Eq. (\ref{Z_Alpha_Final}) can be written for an arbitrary renormalization constant. As a starting point we consider the expression (\ref{LnZM_Pure_Poles}) for $\ln Z(\alpha,\varepsilon^{-1},0)$. It is worth noting that the difference of the left and right hand sides of Eq. (\ref{LnZM_Pure_Poles}) coincides with the renormalization group invariant expression

\begin{equation}\label{Invariant2}
I_2 \equiv \ln\bm{Z}(\bm{\alpha},\varepsilon^{-1}) + \int\limits_{0}^{\bm{\alpha}} d\alpha \frac{\gamma(\alpha)}{\beta(\alpha)} \bigg[\Big(1-\frac{\beta(\alpha)}{\varepsilon\alpha}\Big)^{-1}-1\bigg],
\end{equation}

\noindent
in which $\bm{\alpha}$ is the $D$-dimensional renormalized coupling constant. The independence of $I_2$ from $\mu$ is demonstrated in Appendix \ref{Appendix_Invariants}. Calculating $I_2$ at $\mu=\Lambda$ (when the $D$-dimensional renormalized coupling constant $\bm{\alpha}$ coincides with the four-dimensional renormalized coupling constant $\alpha$) and comparing the result with Eq. (\ref{LnZM_Pure_Poles}) we see that $I_2=0$.

Eq. (\ref{LnZM_Result2}) can be rewritten in the form

\begin{equation}
\ln Z(\alpha,\varepsilon^{-1},\ln\Lambda/\mu) - \int\limits_a^\alpha d\alpha\,\frac{\gamma(\alpha)}{\beta(\alpha)} = \exp\Big(\ln\frac{\Lambda}{\mu}\,\beta(\alpha)\frac{\partial}{\partial\alpha}\Big) g(\alpha),
\end{equation}

\noindent
where

\begin{equation}
g(\alpha) \equiv \ln Z(\alpha,\varepsilon^{-1},0) - \int\limits_a^\alpha d\alpha\,\frac{\gamma(\alpha)}{\beta(\alpha)},
\end{equation}

\noindent
and we omit the other arguments for simplicity. After that, we use the equation

\begin{equation}\label{F_From_G}
\exp\Big(\ln\frac{\Lambda}{\mu}\,\beta(\alpha)\frac{\partial}{\partial\alpha}\Big)f[g(\alpha)] = f\Big[\exp\Big(\ln\frac{\Lambda}{\mu}\,\beta(\alpha)\frac{\partial}{\partial\alpha}\Big) g(\alpha)\Big],
\end{equation}

\noindent
which follows from the identity \cite{Kataev:2024xbl}

\begin{equation}
\exp\Big(\ln\frac{\Lambda}{\mu}\,\beta(\alpha)\frac{\partial}{\partial\alpha}\Big)\Big(f_1(\alpha)\, f_2(\alpha)\Big) = \exp\Big(\ln\frac{\Lambda}{\mu}\,\beta(\alpha)\frac{\partial}{\partial\alpha}\Big)f_1(\alpha)\cdot \exp\Big(\ln\frac{\Lambda}{\mu}\,\beta(\alpha)\frac{\partial}{\partial\alpha}\Big) f_2(\alpha)
\end{equation}

\noindent
derived with the help of the general Leibniz product rule. Choosing $f[g(\alpha)] = \exp(g(\alpha))$ in Eq. (\ref{F_From_G}) we obtain the equation for the renormalization constant $Z$

\begin{eqnarray}\label{Equivalent_Identity}
&& \exp\bigg\{- \smash{\int\limits_a^\alpha} d\alpha\,\frac{\gamma(\alpha)}{\beta(\alpha)} \bigg\}\, Z(\alpha,\varepsilon^{-1},\ln\Lambda/\mu)\nonumber\\
&&\qquad\qquad\qquad\qquad = \exp\Big(\ln\frac{\Lambda}{\mu}\,\beta(\alpha)\frac{\partial}{\partial\alpha}\Big)
\bigg(\exp\bigg\{ - \smash{\int\limits_a^\alpha} d\alpha\,\frac{\gamma(\alpha)}{\beta(\alpha)} \bigg\}\, Z(\alpha,\varepsilon^{-1},0) \bigg).\qquad
\end{eqnarray}

\noindent
To simplify it, one can use Eq. (\ref{Algebraic_Identity}) with the operators

\begin{equation}
A \to \ln\frac{\Lambda}{\mu}\,\beta(\alpha)\frac{\partial}{\partial\alpha};\qquad B \to \int\limits_a^\alpha d\alpha\,\frac{\gamma(\alpha)}{\beta(\alpha)},
\end{equation}

\noindent
which evidently satisfy the required equation $[[A,B],B]=0$. Then Eq. (\ref{Equivalent_Identity}) takes the form

\begin{eqnarray}\label{Z_With_Exponent}
&&\, Z(\alpha,\varepsilon^{-1},\ln\Lambda/\mu) = \exp\Big\{\ln\frac{\Lambda}{\mu}\Big(\beta(\alpha)\frac{\partial}{\partial\alpha} - \gamma(\alpha)\Big)\Big\}\, Z(\alpha,\varepsilon^{-1},0)\nonumber\\
&& \qquad\qquad\qquad\qquad = \exp\Big\{\ln\frac{\Lambda}{\mu}\Big(\beta(\alpha)\frac{\partial}{\partial\alpha} - \gamma(\alpha)\Big)\Big\}\,
\exp\Big\{\int\limits_0^\alpha d\alpha\, \frac{\gamma(\alpha)}{\beta(\alpha)-\varepsilon\alpha}\Big\},\qquad
\end{eqnarray}

\noindent
where we took Eq. (\ref{LnZM_Pure_Poles}) into account. Eq. (\ref{Z_With_Exponent}) evidently has the renormalization group origin because it simply coincides with Eq. (\ref{Z_Rescaling}) derived with the help of the exponential map for the rescaling subgroup of the renormalization group. For the renormalization constant $Z_\alpha$ from Eq. (\ref{Z_With_Exponent}) with the help of Eq. (\ref{Gamma_Alpha}) we immediately obtain the expression (\ref{Z_Alpha_Final}).

As a correctness test, we have verified that Eq. (\ref{Z_With_Exponent}) exactly reproduces the five-loop expression for $\ln Z$ given by Eq. (A.4) in \cite{Meshcheriakov:2023fmk}.

Thus, we have demonstrated that all equations relating coefficients at higher $\varepsilon$-poles, logarithms, and mixed terms in various renormalization constants have the renormalization group origin and follow from Eqs. (\ref{F_With_Exponent}) and (\ref{Z_With_Exponent}).

\section{Conclusion}
\hspace*{\parindent}

As is well known, the coefficients at higher $\varepsilon$-poles \cite{tHooft:1973mfk} and higher logarithms \cite{Collins:1984xc} in the renormalization constants are related to the coefficients of the $\beta$-function and anomalous dimension. However, the dependence of the renormalization constants on $\varepsilon^{-1}$ and on $\ln\Lambda/\mu$ is quite different, although the coefficients at simple $\varepsilon$-poles and (the first power of) $\ln\Lambda/\mu$ differ by a number of loops \cite{Chetyrkin:1980sa}. It is convenient to investigate the relations between the coefficients at higher $\varepsilon$-poles and higher logarithms using a version of the dimensional regularization in which the regularization parameter $\Lambda$ does not coincide with the renormalization point $\mu$. Here for this regularization we rewrite all-order formulas for these coefficients in the MS-like
renormalization schemes in such a way that their renormalization group origin becomes clear. In particular, it is demonstrated that the dependence on $\ln\Lambda/\mu$ can easily be obtained with the help of the exponential map. Consequently, it becomes possible to present the exact dependence of the renormalization constants on $\ln\Lambda/\mu$ (as well as on $\varepsilon^{-1}$) in a visual and beautiful form and construct a simple derivation of the corresponding equations (\ref{Z_Alpha_Final}) and (\ref{Z_With_Exponent}).

\section*{Acknowledgments}
\hspace*{\parindent}

The work of N.M. has been supported by Theoretical Physics and Mathematics Advancement Foundation ``BASIS'', grant No. 21-2-2-25-1.

\appendix

\section{All-loop equations for the renormalization constants in the MS-like schemes}
\hspace*{\parindent}\label{Appendix_Section_Rewriting}

In this appendix we rewrite the explicit all-order expressions for the renormalization constants in the MS-like schemes derived in \cite{Meshcheriakov:2023fmk} in a different form, which reveals their renormalization group origin.

\subsection{Expression for $\ln Z_\alpha$}
\hspace*{\parindent}\label{Appendix_Subsection_LnZ_Alpha}

The expression for $\ln Z_\alpha$, where $Z_\alpha = \alpha/\alpha_0$, in an arbitrary MS-like subtraction scheme is given by Eq. (\ref{LnZ_Original_Result}). Expanding the exponential function in its right hand side into a Taylor series this equation can be presented in the equivalent form

\begin{equation}\label{LnZ_Series}
\frac{\partial\ln Z_\alpha}{\partial\ln\alpha} = 1- \Big(1-\frac{\beta(\alpha)}{\varepsilon\alpha}\Big)^{-1} - \sum\limits_{p=1}^\infty \frac{1}{p!} \ln^p\frac{\Lambda}{\mu}\, \frac{\hat d}{d\ln\alpha} \frac{\beta(\alpha)}{\alpha} \cdot\ldots \cdot \frac{\hat d}{d\ln\alpha} \bigg[\frac{\beta(\alpha)}{\alpha} \Big(1-\frac{\beta(\alpha)}{\varepsilon\alpha}\Big)^{-1}\bigg],
\end{equation}

\noindent
where a number of the $\beta$-functions in each term is equal to $p$. For the expression in the square brackets it is expedient to use the equation

\begin{equation}
\frac{\beta(\alpha)}{\alpha} \Big(1-\frac{\beta(\alpha)}{\varepsilon\alpha}\Big)^{-1} = \beta(\alpha) \frac{d}{d\alpha} \bigg\{\int\limits_0^\alpha \frac{d\alpha}{\alpha}
\bigg[\Big(1-\frac{\beta(\alpha)}{\varepsilon\alpha}\Big)^{-1}-1\bigg] +\ln\alpha\bigg\}.
\end{equation}

\noindent
Indeed, using this identity and taking into account that $Z_\alpha(\alpha=0,\varepsilon^{-1},\ln\Lambda/\mu)=1$ we cast the expression for $\ln Z_\alpha$ in the form\footnote{Note that here (unlike Eq. (\ref{LnZ_Series})) the summation index $p$ starts out equal to 0.}

\begin{eqnarray}
&& \ln Z_\alpha = \ln\alpha -\sum\limits_{p=0}^\infty \frac{1}{p!}\ln^p\frac{\Lambda}{\mu}\, \Big(\beta(\alpha)\frac{d}{d\alpha}\Big)^p \bigg\{\int\limits_0^\alpha \frac{d\alpha}{\alpha}
\bigg[\Big(1-\frac{\beta(\alpha)}{\varepsilon\alpha}\Big)^{-1}-1\bigg] +\ln\alpha\bigg\} \qquad\nonumber\\
&& = \ln\alpha - \exp\Big(\ln\frac{\Lambda}{\mu}\, \beta(\alpha)\frac{\partial}{\partial\alpha} \Big)\,\bigg\{\int\limits_0^\alpha \frac{d\alpha}{\alpha}
\bigg[\Big(1-\frac{\beta(\alpha)}{\varepsilon\alpha}\Big)^{-1}-1\bigg] +\ln\alpha\bigg\}.
\end{eqnarray}

\noindent
This equation can equivalently be rewritten as

\begin{equation}
\ln \alpha_0 = \ln(\alpha Z_\alpha^{-1}) = \exp\Big(\ln\frac{\Lambda}{\mu}\,\beta(\alpha)\frac{\partial}{\partial\alpha}\Big)\,
\bigg\{\int\limits_0^\alpha \frac{d\alpha}{\alpha}
\bigg[\Big(1-\frac{\beta(\alpha)}{\varepsilon\alpha}\Big)^{-1}-1\bigg] +\ln\alpha\bigg\}
\end{equation}

\noindent
and coincides with Eq. (\ref{F_With_Exponent}) for the particular case $f(\alpha_0)=\ln\alpha_0$.

\subsection{Expression for $(Z_\alpha)^S$}
\hspace*{\parindent}\label{Appendix_Subsection_Z_Alpha_S}

Similarly, the all-loop expressions for $(Z_\alpha)^S$ in an arbitrary MS-like scheme is given by Eq. (\ref{ZS_Original_Result}). Again, expanding the right hand side into a Taylor series we obtain

\begin{eqnarray}\label{ZS_Series}
&&\hspace*{-5mm} \Big(\frac{\partial}{\partial\ln\alpha} - S\Big) (Z_\alpha)^S = \bigg[- S\sum\limits_{p=1}^\infty \frac{1}{p!} \ln^p\frac{\Lambda}{\mu}\, \Big(\frac{\hat d}{d\ln\alpha}-S\Big) \frac{\beta(\alpha)}{\alpha} \Big(\frac{\hat d}{d\ln\alpha}-S\Big) \frac{\beta(\alpha)}{\alpha} \cdot \ldots  \nonumber\\
&&\hspace*{-5mm}\qquad\qquad\qquad\qquad\qquad\qquad\quad \times \Big(\frac{\hat d}{d\ln\alpha}-S\Big) \frac{\beta(\alpha)}{\alpha} - S\bigg] \Big(1-\frac{\beta(\alpha)}{\varepsilon\alpha }+ S \int\limits^{\wedge} \frac{d\alpha}{\alpha}\, \frac{\beta(\alpha)}{\varepsilon\alpha} \Big)^{-1},\qquad
\end{eqnarray}

\noindent
where each term in the square brackets contains $p$ factors $\beta(\alpha)$. The rightmost expression here can be presented in the form

\begin{eqnarray}\label{Operator_Equation}
&& \Big(1-\frac{\beta(\alpha)}{\varepsilon\alpha }+ S \int\limits^{\wedge} \frac{d\alpha}{\alpha}\, \frac{\beta(\alpha)}{\varepsilon\alpha} \Big)^{-1}
= \Big(1-\frac{\beta(\alpha)}{\varepsilon\alpha}\Big)^{-1} \exp\bigg\{-S\int\limits_0^\alpha\frac{d\alpha}{\alpha}\bigg[\Big(1-\frac{\beta(\alpha)}{\varepsilon\alpha}\Big)^{-1}-1\bigg]\bigg\}\nonumber\\
&& = - S^{-1} \Big(\frac{d}{d\ln\alpha} - S\Big) \exp\bigg\{-S\int\limits_0^\alpha\frac{d\alpha}{\alpha}\bigg[\Big(1-\frac{\beta(\alpha)}{\varepsilon\alpha}\Big)^{-1}-1\bigg]\bigg\}
\end{eqnarray}

\noindent
because

\begin{eqnarray}
&& \Big(1-\frac{\beta(\alpha)}{\varepsilon\alpha }+ S \int\limits^{\wedge} \frac{d\alpha}{\alpha}\, \frac{\beta(\alpha)}{\varepsilon\alpha} \Big) \Big(1-\frac{\beta(\alpha)}{\varepsilon\alpha}\Big)^{-1} \exp\bigg\{-S\int\limits_0^\alpha\frac{d\alpha}{\alpha}\bigg[\Big(1-\frac{\beta(\alpha)}{\varepsilon\alpha}\Big)^{-1}-1\bigg]\bigg\} \qquad\nonumber\\
&& = \Big(1 -\int\limits_0^\alpha d\alpha\, \frac{d}{d\alpha}\Big)\exp\bigg\{-S\int\limits_0^\alpha\frac{d\alpha}{\alpha}\bigg[\Big(1-\frac{\beta(\alpha)}{\varepsilon\alpha}\Big)^{-1}-1\bigg]\bigg\} = 1.
\end{eqnarray}

\noindent
(Note that Eq. (\ref{Operator_Equation}) establishes the correspondence between Eqs. (\ref{LnZ_Original_Result}) and (\ref{ZS_Original_Result}) for $\ln Z_\alpha$ and $(Z_\alpha)^S$, respectively, for the case $\Lambda=\mu$.)

From Eq. (\ref{Operator_Equation}) we see that both sides of Eq. (\ref{ZS_Series}) contain the operator $\partial/\partial\ln\alpha - S$, which can be removed (taking into account the boundary condition $Z_\alpha(\alpha=0,\varepsilon^{-1},\ln\Lambda/\mu)=1$). Then we obtain

\begin{eqnarray}
&&\hspace*{-5mm} (Z_\alpha)^S = \bigg\{1 +\sum\limits_{p=1}^\infty \frac{1}{p!} \ln^p\frac{\Lambda}{\mu}\, \frac{\beta(\alpha)}{\alpha} \Big(\frac{\hat d}{d\ln\alpha}-S\Big) \cdot \ldots \cdot \frac{\beta(\alpha)}{\alpha} \Big(\frac{\hat d}{d\ln\alpha}-S\Big) \bigg\}\nonumber\\
&&\hspace*{-5mm}\qquad\qquad\qquad\qquad\qquad\qquad\qquad\qquad\qquad\quad \times\exp\bigg\{-S\int\limits_0^\alpha\frac{d\alpha}{\alpha}\bigg[\Big(1-\frac{\beta(\alpha)}{\varepsilon\alpha}\Big)^{-1}-1\bigg]\bigg\}.\qquad
\end{eqnarray}

\noindent
It is expedient to insert the factor $1=\alpha^S\alpha^{-S}$ before the exponent and commute $\alpha^S$ to the left using the identity

\begin{equation}
\Big(\alpha \frac{\hat d}{d\alpha} - S\Big) \alpha^S = \alpha^S \cdot \alpha \frac{\hat d}{d\alpha},
\end{equation}

\noindent
where hats stress that the derivatives in particular act on expressions on the right that were not explicitly written out. After that, the result can be presented in the form (\ref{Arbitrary_F}),

\begin{equation}
\alpha_0^{-S} = \alpha^{-S} (Z_\alpha)^S = \exp\Big(\ln\frac{\Lambda}{\mu}\,\beta(\alpha) \frac{\partial}{\partial\alpha}\Big)\,\alpha^{-S} \exp\bigg\{-S\int\limits_0^\alpha\frac{d\alpha}{\alpha}\bigg[\Big(1-\frac{\beta(\alpha)}{\varepsilon\alpha}\Big)^{-1}-1\bigg]\bigg\},
\end{equation}

\noindent
where $f(\alpha_0) = \alpha_0^{-S}$.

\subsection{Expression for $\ln Z$}
\hspace*{\parindent}\label{Appendix_Subsection_LnZ}

For the field renormalization constant $Z$ the all-loop expression relating the coefficients at higher $\varepsilon$-poles and logarithms to the coefficients of the $\beta$-function and anomalous dimension is given by Eq. (\ref{LnZM_Original_Result}). As earlier, we expand the exponential the right hand side into a Taylor series,

\begin{eqnarray}\label{LnZM_Series}
&& \frac{\partial\ln Z}{\partial\ln\alpha} = \frac{\alpha\gamma(\alpha)}{\beta(\alpha)}\Big[1 - \Big(1-\frac{\beta(\alpha)}{\varepsilon\alpha}\Big)^{-1}\Big]\nonumber\\
&&\qquad\qquad -\sum\limits_{p=1}^\infty \frac{1}{p!} \ln^p\frac{\Lambda}{\mu}\, \frac{\hat d}{d\ln\alpha} \frac{\beta(\alpha)}{\alpha} \frac{\hat d}{d\ln\alpha}
\cdot \ldots \cdot \frac{\beta(\alpha)}{\alpha} \frac{\hat d}{d\ln\alpha} \Big[\gamma(\alpha)\Big(1-\frac{\beta(\alpha)}{\varepsilon\alpha}\Big)^{-1}\Big],\qquad
\end{eqnarray}

\noindent
where each term contains $p$ derivatives with respect to $\ln\alpha$, and use the identity

\begin{equation}
\gamma(\alpha) \Big(1-\frac{\beta(\alpha)}{\varepsilon\alpha}\Big)^{-1} = \beta(\alpha) \frac{d}{d\alpha} \int\limits_0^\alpha d\alpha\, \frac{\gamma(\alpha)}{\beta(\alpha)} \Big[\Big(1-\frac{\beta(\alpha)}{\varepsilon\alpha}\Big)^{-1}-1\Big] + \beta(\alpha) \frac{d}{d\alpha} \int\limits_a^\alpha d\alpha \frac{\gamma(\alpha)}{\beta(\alpha)}.
\end{equation}

\noindent
(In the last integral the constant $a$ at the lower limit of integration can be arbitrary.) Then, removing the derivative with respect to $\ln\alpha$ (certainly, using the evident condition 
$\ln Z(\alpha=0,\varepsilon^{-1},\ln\Lambda/\mu)=0$) we cast the expression for $\ln Z$ in the form

\begin{eqnarray}\label{LnZM_Preliminary}
&& \ln Z = -\exp\Big(\ln\frac{\Lambda}{\mu}\,\beta(\alpha)\frac{\partial}{\partial\alpha}\Big)  \int\limits_0^\alpha d\alpha\, \frac{\gamma(\alpha)}{\beta(\alpha)} \Big[\Big(1-\frac{\beta(\alpha)}{\varepsilon\alpha}\Big)^{-1}-1\Big]
\nonumber\\
&&\qquad\qquad\qquad\qquad\qquad\qquad\qquad + \Big[1-\exp\Big(\ln\frac{\Lambda}{\mu}\,\beta(\alpha)\frac{\partial}{\partial\alpha}\Big)\Big]\, \int\limits_a^\alpha d\alpha\,\frac{\gamma(\alpha)}{\beta(\alpha)}.\qquad
\end{eqnarray}

\noindent
Taking into account that the renormalization constants in $D$ and 4 dimensions for the considered regularization are related as $\bm{Z}(\alpha,\varepsilon^{-1}) = Z(\alpha,\varepsilon^{-1},\ln\Lambda/\mu=0)$ (see, e.g., \cite{Meshcheriakov:2023fmk}) from Eqs. (\ref{RGFs_D_Definition}) and (\ref{RGFs_Relations}) we obtain

\begin{equation}
\gamma(\alpha) = \Big(\beta(\alpha)-\varepsilon\alpha\Big) \frac{\partial\ln \bm{Z}(\alpha,\varepsilon^{-1})}{\partial\alpha} = \Big(\beta(\alpha)-\varepsilon\alpha\Big) \frac{\partial\ln Z(\alpha,\varepsilon^{-1},0)}{\partial\alpha},
\end{equation}

\noindent
so that

\begin{equation}
\ln Z(\alpha,\varepsilon^{-1},0) = \int\limits_0^\alpha d\alpha\, \frac{\gamma(\alpha)}{\beta(\alpha)-\varepsilon\alpha} = - \int\limits_0^\alpha d\alpha\, \frac{\gamma(\alpha)}{\beta(\alpha)} \Big[\Big(1-\frac{\beta(\alpha)}{\varepsilon\alpha}\Big)^{-1}-1\Big].
\end{equation}

\noindent
This implies that Eq. (\ref{LnZM_Preliminary}) can equivalently be rewritten in the form (\ref{LnZM_Result2}),

\begin{equation}
\ln Z(\alpha,\varepsilon^{-1},\ln\Lambda/\mu) - \int\limits_a^\alpha d\alpha\,\frac{\gamma(\alpha)}{\beta(\alpha)} = \exp\Big(\ln\frac{\Lambda}{\mu}\,\beta(\alpha)\frac{\partial}{\partial\alpha}\Big)
\bigg[ \ln Z(\alpha,\varepsilon^{-1},0) - \int\limits_a^\alpha d\alpha\,\frac{\gamma(\alpha)}{\beta(\alpha)} \bigg].
\end{equation}

\section{Renormalization group invariance of the expressions (\ref{Invariant1}) and (\ref{Invariant2})}
\hspace*{\parindent}\label{Appendix_Invariants}

Differentiating the expression (\ref{Invariant1}) with respect to $\ln\mu$ and taking into account Eqs. (\ref{RGFs_D_Definition}) and (\ref{RGFs_Relations}) we obtain

\begin{eqnarray}
&& \frac{d}{d\ln\mu} I_1\bigg|_{\alpha_0=\text{const}} = \frac{d}{d\ln\mu}\Bigg(\Big(\frac{\mu}{\Lambda}\Big)^\varepsilon \bm{\alpha}\exp\bigg\{\int\limits_0^{\bm{\alpha}}\frac{d\alpha}{\alpha}\bigg[\Big(1-\frac{\beta(\alpha)}{\varepsilon\alpha}\Big)^{-1}-1\bigg]\bigg\}\Bigg) \nonumber\\
&&\qquad\qquad\qquad\qquad = \frac{I_1}{\bm{\alpha}} \bigg\{\varepsilon\bm{\alpha} + \beta(\bm{\alpha}) - \varepsilon\bm{\alpha} + \Big(\beta(\bm{\alpha}) - \varepsilon\bm{\alpha}\Big)\Big(\frac{\varepsilon\bm{\alpha}}{\varepsilon\bm{\alpha} - \beta(\bm{\alpha})} -1\Big)\bigg\} = 0.\qquad
\end{eqnarray}

\noindent
(Note that it is the $D$-dimensional renormalized coupling $\bm{\alpha}$ that enters this expression.)

Similarly, the derivative of the expression $I_2$ defined by Eq. (\ref{Invariant2}) with respect to $\ln\mu$ also vanishes,

\begin{equation}
\frac{d}{d\ln\mu} I_2\bigg|_{\alpha_0=\text{const}} = \gamma(\bm{\alpha}) + \Big(\beta(\bm{\alpha}) - \varepsilon\bm{\alpha}\Big)
\frac{\gamma(\bm{\alpha})}{\beta(\bm{\alpha})} \Big(\frac{\varepsilon\bm{\alpha}}{\varepsilon\bm{\alpha} - \beta(\bm{\alpha})}-1\Big) = 0.
\end{equation}

\end{document}